\documentclass[aps,prb,twocolumn,showpacs,showkeys, superscriptaddress,preprintnumbers,amsmath,amssymb]{revtex4}

\usepackage{txfonts}
\usepackage{amssymb}

\usepackage{graphicx}
\usepackage{dcolumn}
\usepackage{bm}
\usepackage{color}

\begin{document}


\title{Topologically trivial and nontrivial edge bands in graphene induced by irradiation}

\author{Mou Yang}
\altaffiliation{Electronic address: yang.mou@hotmail.com}

\author{Zhi-Jun Cai}
\author{Rui-Qiang Wang}

\affiliation{Guangdong Provincial Key Laboratory of Quantum Engineering and Quantum Materials, \\
School of Physics and Telecommunication Engineering,
South China Normal University, Guangzhou 510006, China}

\author{Yan-Kui Bai}
\affiliation{College of Physical Science and Information Engineering and Hebei Advance Thin Films Laboratory, \\ Hebei Normal University, Shijiazhuang, Hebei 050024, China}

\begin{abstract}

We proposed a minimal model to describe the Floquet band structure of two-dimensional materials with light-induced resonant inter-band transition. We applied it to graphene to study the band features caused by the light irradiation. Linearly polarized light induces pseudo gaps (gaps are functions of wavevector), and circularly polarized light causes real gaps on the quasi-energy spectrum. If the polarization of light is linear and along the longitudinal direction of zigzag ribbons, flat edge bands appear in the pseudo gaps, and if is in the lateral direction of armchair ribbons, curved edge bands can be found. For the circularly polarized cases, edge bands arise and intersect in the gaps of both types of ribbons. The edge bands induced by the circularly polarized light are helical and those by linearly polarized light are topologically trivial ones. The Chern number of the Floquet band, which reflects the number of pairs of helical edge bands in graphene ribbons, can be reduced into the winding number at resonance.

\end{abstract}


\maketitle

\section{Introduction}

%

%
%

Graphene has drawn much attention since it was discovered in the laboratory.\cite{graphene,graphene_review} Graphene has a number of interesting physical properties and has a great potential for application. Pristine graphene is a gapless Dirac material, while the energy gap is needed for the fabrication of switching devices. There are a few causes, such as staggered substrate influence\cite{stagger_1,stagger_2,stagger_3} and the spin-orbit coupling,\cite{SOC_1,SOC_2,SOC_3} to open a gap on the spectrum of graphene. The latter is more attractive because it makes graphene a topologic insulator and leads to helical edge bands which is topologically protected by the time-reversal symmetry.\cite{SOC_1} However, the spin-orbit coupling in graphene is proven to be too weak to detect.\cite{graphene_review} Recent researches implied that the time dependent driving may have the similar effects as the spin-orbit coupling in graphene: it generates gaps and turn a normal material into a special topologic insulator called Floquet topologic insulator.\cite{FTI_1,FTI_2,FTI_3,FTI_4,FTI_5} Besides in condensed matters, the interest of the novel effects of driving is increasing in cold atoms\cite{cold_atom_1,cold_atom_2,cold_atom_3,cold_atom_4} and other fields. Recently, the Floquet topologic phase was realized in a photonic crystal,\cite{exp_photonic} which indicated the validity of the theoretical prediction. Light irradiation is an important periodically driving source, and the irradiation induced energy gaps in a topological insulator was observed recently.\cite{exp_TI} These experiments provide  the probability to generate gaps and change the topologic property of graphene by light irradiation.\cite{graphene_irradiated_1,graphene_irradiated_2,graphene_irradiated_3,graphene_irradiated_4}

Light irradiation generates energy gaps in graphene by two mechanisms. First, under the affection of light, the electron near the Dirac point emits a photon and re-absorbs it to renormalize the band structure, and a gap is generated at the Dirac point to separate the conduction and valence bands,\cite{nonresonant_1,nonresonant_2,nonresonant_3,nonresonant_4} which is the effect of second order perturbation. Second, light induces resonant transition between conduction band and valence band states, and produces dynamic gaps on the quasi-energy spectrum at $E=\pm\hbar\omega/2$, where $\omega$ is the angular frequency of light.\cite{resonant_1,resonant_2,resonant_3,resonant_4} The latter is more attractive because it is a first order process.

Typically, periodically driven system is treated in frequency space,\cite{Floquet_1,Floquet_2,Floquet_3} also called Floquet space. The whole space is divided into infinite subspaces according to the number of photons absorbed and emitted. The system is solved by truncating the Floquet space at a finite dimension. For the weak driven cases, the main physics is determined by the one-photon processes that can be well understood. It is possible to develop a short theory to handle the driven system by only taking the one-photon processes into account. The theory should simplify the calculation, reproduce the results of other more complicated methods, and more importantly, give more insight on the physics of driven systems.

In this paper, we proposed a minimal model to describe the Floquet band structure of two-dimensional materials with light-induced resonant inter-band transition and applied the theory to graphene. Linearly polarized light induced pseudo gaps, and circular polarized light causes real gaps on the Floquet quasi-energy spectrum of graphene. For the circular polarization cases, edge bands arise in the gaps and intersect for both zigzag and armchair ribbons. Interestingly, linear polarized illumination can also lead to edge bands, depending on the type of ribbon and the polarization orientation. If the polarization is longitudinal along zigzag ribbons, flat edge bands appear in the pseudo gaps, and if in the lateral direction of armchair ribbons, curved edge bands arise. The topologic property of the Floquet bands is reflected by the Chern number, and we found it can be reduced into the winding number at resonance. The edge bands induced by the circular polarized light are helical and those by linear polarized light are topologically trivial ones. 

\section{Floguet theory of inter-band optic transition}

We consider a two-band system consisting of one conduction band and one valence band. When a laser normally irradiates on the graphene sheet, a time-dependent vector potential $\bm{\mathcal{A}}(t)= \bm{A}\cos\omega t$ is introduced, where $\boldsymbol{A}$ is the amplitude vector of $\bm{\mathcal{A}}$, $\omega$ is the angular frequency, and $t$ is the time. If the system is weakly perturbed, in the frame of  $A\cdot p$  approximation, the time-dependent Hamiltonian reads
\begin{eqnarray}\label{Ht_original}
H_0(t) &=& H_{\bm k}  - \bm{\mathcal{A}}(t) \cdot \bm{p},
\end{eqnarray}
where $H_{\bm k}$ is the Hamiltonian without light irradiating, and $\boldsymbol{p}$ is the momentum operator. The eigen values of $H_{\bm k}$ are the conduction and valence band energies denoted by $\epsilon_c$ and $\epsilon_v$, and the corresponding eigen states are $|c\rangle$ and $|v\rangle$, respectively. In Eq. (\ref{Ht_original}), the electron charge $e$ and the electron effective mass $m$ are set to be 1. In basis of $|c\rangle$ and $|v\rangle$, the time-dependent Hamiltonian (rotating wave approximate is used) can be written as
\begin{eqnarray}\label{Ht}
H(t)=
\left(\begin{array}{*{20}c}
\epsilon_c & \frac 12 g^*e^{ -i\omega t} \\
\frac 12 ge^{ i\omega t} & \epsilon_v
\end{array}\right),
\end{eqnarray}
where $g$ is the transition element defined by
\begin{eqnarray}\label{g}
g = -\boldsymbol{A} \cdot\langle v| \, \boldsymbol{p} \, | c \rangle.
\end{eqnarray}
The time-dependent Schr\"odinger equation $i\partial_t \psi = H(t)\psi$, in which the $\hbar$ is set to be 1, can be reduced into a static one $\mathcal{H}\psi=E\psi$ by introducing the unitary transformation
\begin{eqnarray}\label{Ut}
U=
\left(\begin{array}{*{20}c}
e^{-i(\epsilon_c-\delta/2)t} & 0 \\
0 & e^{-i(\epsilon_v+\delta/2)t}
\end{array}\right),
\end{eqnarray}
where $\delta = (\epsilon_c-\epsilon_v) -\omega$ is the detune. The static Hamiltonian is obtained by
\begin{eqnarray}\label{H}
\mathcal{H} = UHU^+ + iU \frac{\partial U^+}{\partial t} = \frac12 \left(\begin{array}{*{20}c} 
\delta & g^* \\
 g & -\delta
\end{array}\right).
\end{eqnarray}
Solving the eigen problem of the static Hamiltonian, we have the eigen pairs
\begin{eqnarray}\label{eigen}
E_\pm = \pm \frac 12 D, \quad \psi_\pm= \frac{1}{\sqrt{2}}
\left(\begin{array}{*{20}c} 
\sqrt{|\pm1+\delta/D|} \\
\pm e^{i\theta}\sqrt{|\pm1-\delta/D|}
\end{array}\right),
\end{eqnarray}
where $D =\sqrt{|g|^2+\delta^2}$ and $\theta={\rm arg}(g)$ is the complex angle. Go back to the basis of $|c\rangle$ and $|v\rangle$, and we have the quantum states satisfying the time-dependent Shr\"{o}dinger equation for $H(t)$, 
\begin{eqnarray}
U\psi_\pm e^{-iE_\pm t}.
\end{eqnarray}

According to the Floquet theorem, the solutions of time-dependent Shr\"{o}dinger equation for periodic time-dependent Hamiltonian  must be of the form $\psi = e^{-iE^Ft} \psi^F$, where $E^F$ is time independent and $\psi^F$ is of the same period as $H(t)$. The quantities $E^F$ and $\psi^F$ are called as Floquet energy and Floquet state respectively, which are the solution pair of the Floquet equation $H^F\psi^F = E^F \psi^F$, where $H^F = H(t)-i\partial_t$ is the Floquet operator. One can verify that, if $E^F$ and $\psi^F$ satisfy the Floquet equation, $E^F+n\omega$ and $\psi^Fe^{in\omega t}$ for arbitrary integer $n$ are also a Floquet pair. To eliminate the non-uniqueness, we choose proper $n$ so that the Floquet energies are recovered to the the conduction and valence band energies for infinitesimal weak driven intensity. After doing so, we have the Floquet energies
\begin{eqnarray}\label{EF}
E^F_\pm = \frac12 (\epsilon_c + \epsilon_v) \pm \frac12 \left( \eta D+\omega \right),
\end{eqnarray}
where $\eta$ is the sign of $\delta$. The corresponding Floquet states are
\begin{eqnarray}\label{psiF}
\begin{split}
\psi_+^F &= a_F|c\rangle + b_F|v\rangle e^{i\omega t},\\
\psi_-^F &=  b_F^*|c\rangle e^{-i\omega t} - a_F|v\rangle,
\end{split}
\end{eqnarray}
where the coefficients $a_F$ and $b_F$ are defined as
\begin{eqnarray}\label{ab}
\begin{split}
a_F = \frac1{\sqrt2}\sqrt{1 + \left|\frac{\delta}{D} 
\right|},\quad 
b_F = e^{i\theta}\frac{\eta}{\sqrt2}\sqrt{1 - \left|\frac{\delta}{D} 
\right|}.
\end{split}
\end{eqnarray}
In Eqs. (\ref{EF}) and (\ref{psiF}), when we set $g\rightarrow 0$, the Floquet energies $E_{\pm}^F$ are reduced into $\epsilon_c$ and $\epsilon_v$, and Floquet states $\psi_{\pm}^F$ are recovered to $|c\rangle$ and $|v\rangle$, respectively.

The above derivations are based on linearly polarized irradiation, but also valid for circular polarization by regarding the vector potential amplitude $\bm A$ as a complex quantity. The irradiation has two known effects. (1) It generates resonant gaps on the Floquet spectrum. (2) It can change the topologic property of band structure and create new edge bands, depending on the polarization of irradiation. In the following, we will apply the above theory to graphene and investigate the how the two effects act on graphene.  

\begin{figure}
\includegraphics[width=7cm]{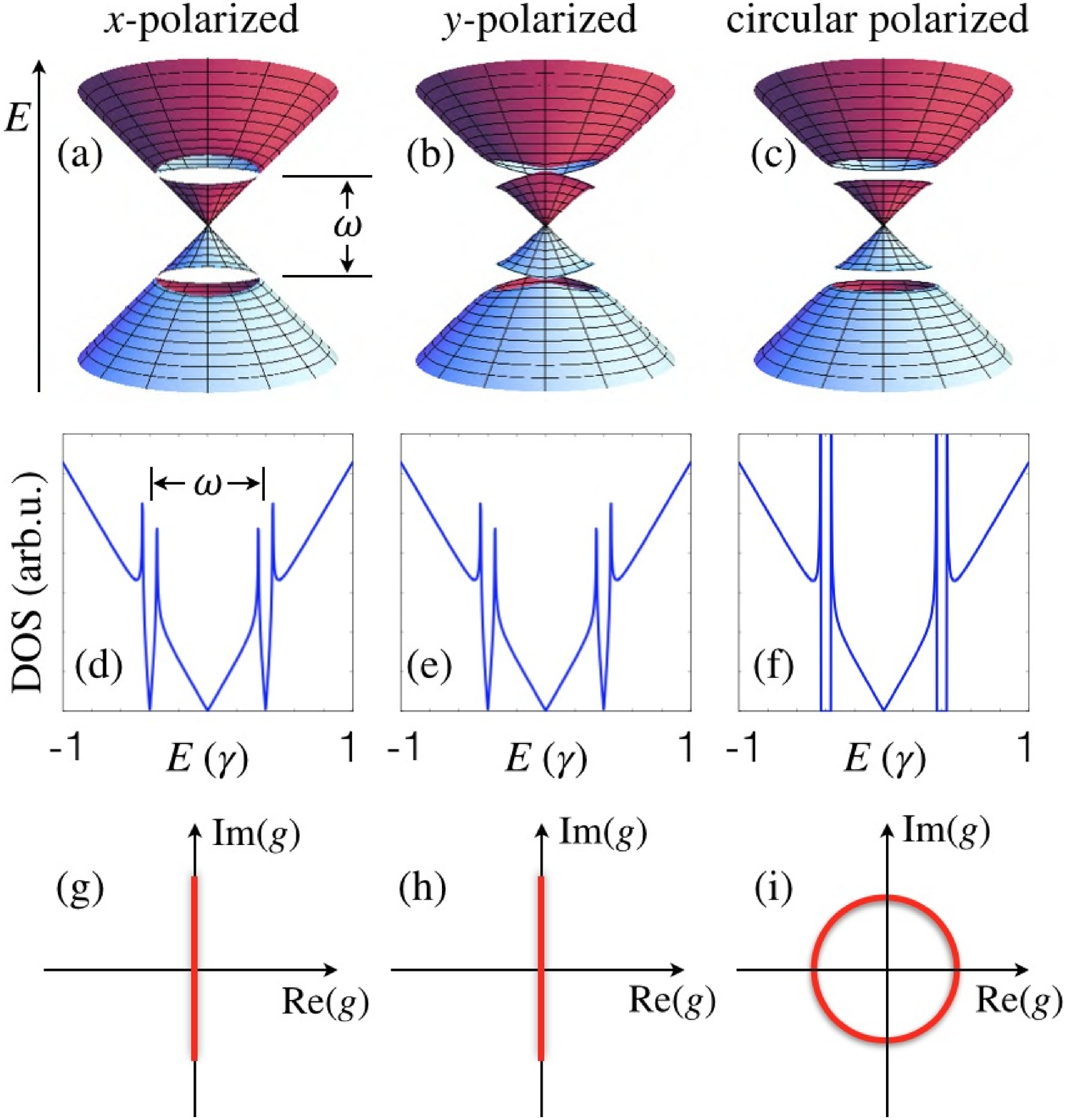}
\caption{(Color Online) \label{dos}
(a-c) Floguet bands, (d-f) density of states, and (g-i) winding number for bulk graphene sheet under $x$-polarized irradiation with $A_x=0.1\gamma$, $y$-polarized irradiation with $A_y=0.1\gamma$, and circularly polarized irradiation with $\sqrt{|A_x|^2+|A_y|^2}=0.1\gamma$. The frequency of the irradiation is $\omega=0.8\gamma$.}
\end{figure}

\section{Resonant gaps of bulk graphene}

There are two non-equivalent valleys in graphene. The low-energy Hamiltonian of valley $K$ reads
\begin{eqnarray}\label{HK}
H_{\boldsymbol{k}} = \boldsymbol{\sigma} \cdot \boldsymbol{k} = 
\left(\begin{array}{*{20}c}
0 & k_x-ik_y \\
 k_x+ik_y & 0
\end{array}\right),
\end{eqnarray}
where $\boldsymbol{\sigma}=(\sigma_x,\sigma_y)$ is the Pauli matrix set, $\boldsymbol{k}=(k_x,k_y)$ is the wavevector, and the Fermi velocity $v_F$ is set to be 1. The band energies and band states are
\begin{eqnarray}\label{eigenK}
\epsilon_{c/v} = \pm k, \quad |c,v\rangle =
\frac1{\sqrt2} \left(\begin{array}{*{20}c}
1 \\
\pm e^{i\varphi}
\end{array}\right),
\end{eqnarray}
where $k=(k_x^2+k_y^2)^{1/2}$ and $\varphi={\rm arg}(k_x+ik_y)$ reflect the amplitude and orientation of $\bm{k}$. When the graphene is under illumination, the Peierls substitution 
$\boldsymbol{k} \rightarrow \boldsymbol{k}-\boldsymbol{\mathcal{A}}$ should be applied, and this leads to 
$H_{\boldsymbol{k}}$ is replaced with $H_{\boldsymbol{k}} -\boldsymbol{\sigma} \cdot \boldsymbol{\mathcal{A}}$.  
Because the momentum is defined by
$\boldsymbol{p} = \nabla_{\boldsymbol{k}} H_{\boldsymbol{k}} =\boldsymbol{\sigma}$,
we have substitution Hamiltonian
$H_{\boldsymbol{k}} - \boldsymbol{\mathcal{A}}\cdot \boldsymbol{p} $, which is just the Hamiltonian in Eq. (\ref{Ht_original}).\cite{explain} So, the $A\cdot P$ approximation can also be applied to graphene\cite{explain}, and the detailed discussion about this can be found in the Appendix.
According to Eq. (\ref{g}), the transition element is calculated as
\begin{eqnarray}\label{gK}
g = i\; (A_x \sin \varphi - A_y \cos \varphi),
\end{eqnarray}
where $A_x$ and $A_y$ are the components of $\boldsymbol{A}$ in $x$- and $y$-directions. Substituting $\epsilon_c$, $\epsilon_v$, and $g$ into Eq. (\ref{EF}), we have the Floquet energies of graphene. On the Floquet energy spectrum, gaps can be found at resonance occurs, saying, $\delta=0$ or equivalently $k=\omega/2$. When $k$ takes the two infinite closed values $\omega/2+0^+$ and $\omega/2-0^+$, $E^F_+$ has two finitely different values, and the difference between them is the gap. By taking $\delta=0^+$ and $0^-$, the gap is calculated as
\begin{eqnarray}\label{gap}
\Delta = |g(\varphi)|.
\end{eqnarray}
As happens to $E^F_+$, another identical gap can be found in the spectrum of $E^F_-$.

If the light is $x$-polarized, we have the gap profile $\Delta= |A_x \sin \varphi|$. Figure \ref{dos} (a) shows the Floquet energy band for this case. One can find the two pseudogaps at $E=\pm \omega/2$. The gaps reach its maximum at $\varphi=\pi/2$ and are closed at $\varphi=0$. If the light is $y$-polarized, we have $\Delta= |A_y \cos \varphi|$. The gaps have maximum and zeros at $\varphi=0$ and $\varphi=\pi/2$, respectively, as shown in Figure \ref{dos} (b). For these linear polarization cases, no real gap opens on the density of states of the Floquet energy bands, as Fig. \ref{dos} (d) and (e) demonstrate.
The gapless spectrum is the consequence of time reversal symmetry of the Hamiltonian in Eq. (\ref{H}) under linearly polarized illumination. The diagonal elements, $\delta$ and $-\delta$, are apparently time-reversal invariant, and the time-symmetry is determined by $g$, the expression of which is given in Eq. (\ref{gK}). The time reversal operation makes $i$ into $-i$ and converts $\bm k$ into $-\bm k$. The latter takes $\varphi$ into $\varphi+\pi$, so $g$ is time-invariant and the time-reversal-symmetry of the system is preserved. 

If the light is circularly polarized, the vector potential amplitude $\bm A$ has to be complex, which can be modeled by $A_x=A$ and $A_y=iA$. According to Eq. (\ref{gK}) and (\ref{gap}), we immediately have $g= Ae^{i\varphi}$ and $\Delta= A$. The gaps are independent of $\varphi$ and are real gaps. Because $\bm A$ is complex, the time-reverse of $\bm A$ is ${\bm A}^*$, the time-reversal counterpart of $g$ is not itself, and the time-reversal symmetry of the system is broken. Figures \ref{dos} (c) and (f) show the Floquet band and the density of states. For ellipse circular polarization, the real gap (the minimum of the gap function $\Delta$) is determined by the vector component along the short radius of polarization ellipse, saying, 
\begin{eqnarray}\label{gap_min}
\Delta_{\rm min} = |A_{\rm short}|.
\end{eqnarray}
Letting the short radius to be zero, the case is reduced to the linear polarization. 

\section{Electronic structures of graphene ribbons under laser irradiation}

\begin{figure}
\includegraphics[width=8cm]{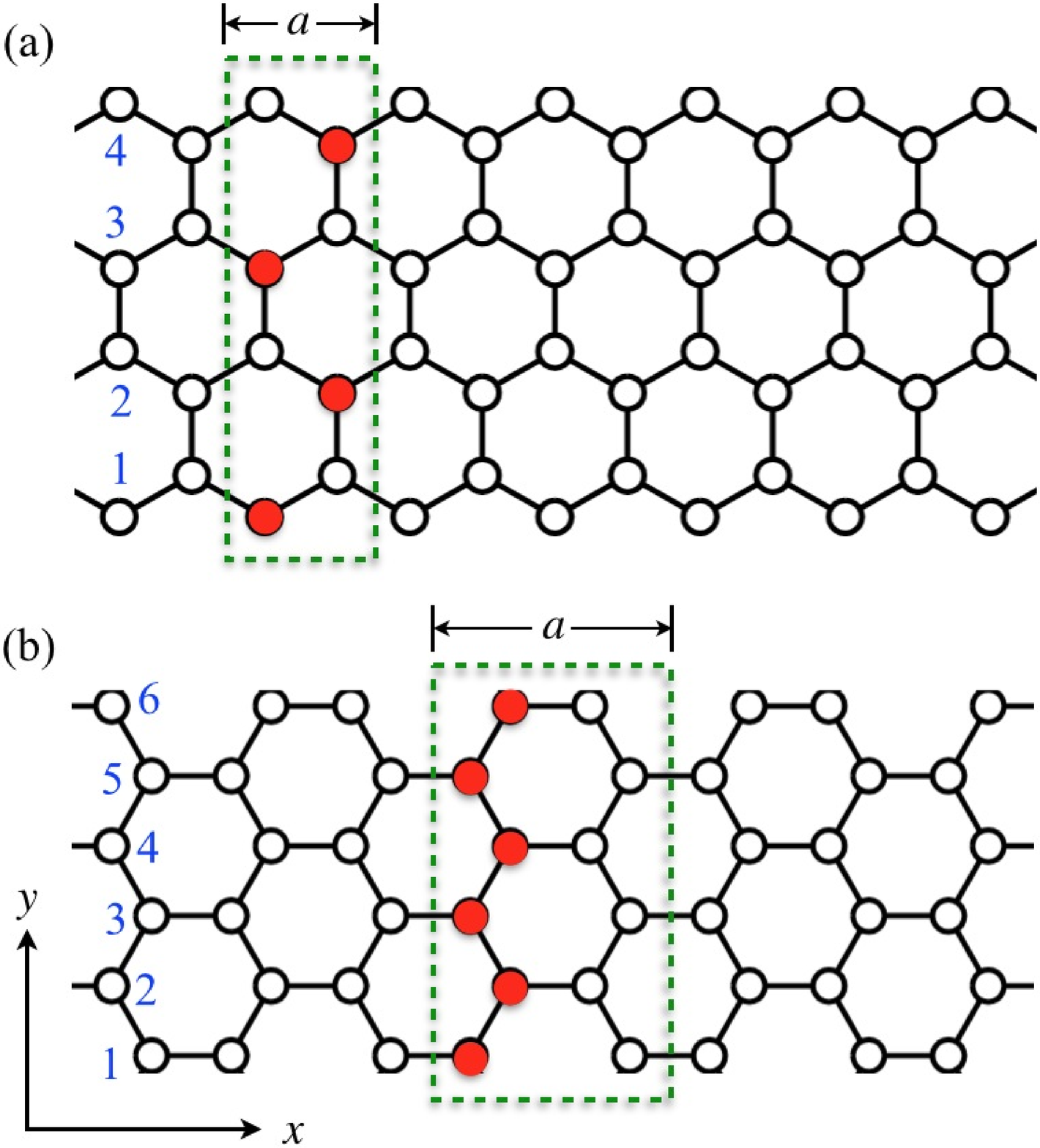}
\caption{(Color Online) \label{ribbon}
Sketches of (a) zigzag and (b) armchair graphene ribbons. The rectangles illustrate the translational unit cells. The wavefunction on the atoms represented by the filled circles will be studied.}
\end{figure}

\begin{figure}
\includegraphics[width=8cm]{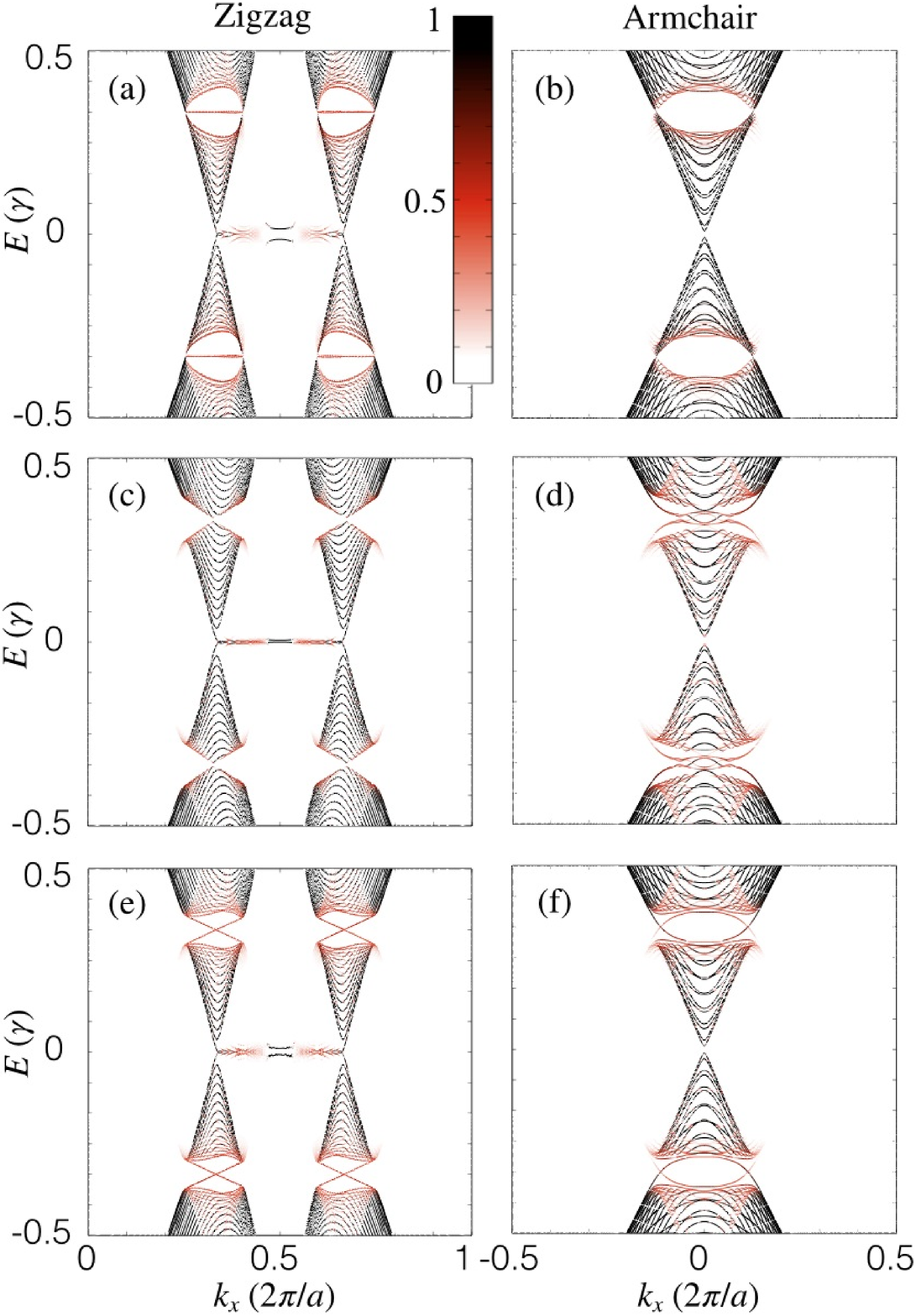}
\caption{(Color Online) \label{dispersion}
Weighted dispersions of graphene ribbons with 100 lateral atoms under  (a-b) $x$-polarized irradiation with $A_x=0.1\gamma$, (c-d) $y$-polarized irradiation with $A_y=0.1\gamma$, and (e-f) circularly polarized irradiation with $\sqrt{|A_x|^2+|A_y|^2}=0.1\gamma$. The frequency of the irradiation is $\omega=0.8\gamma$ and the labeling rule of atoms in lateral direction can be found in Fig. \ref{ribbon}.}
\end{figure}

We use tight-binding model to investigate the electronic structures of graphene ribbons under irradiation to check the existence of edge bands in the resonant gaps. The two typical types of graphene ribbons, the zigzag and armchair ones, and their layout, are demonstrated in Figure \ref{ribbon}. The tight-binding Hamiltonian is
\begin{eqnarray}\label{Htb}
H_{\rm tb} = \gamma \sum_{\langle i,j \rangle}  c^+_ic_j,
\end{eqnarray}
where $\gamma=2.7$eV is the carbon-carbon hopping energy, $c^+_i$ $(c_i)$ is the creation (annihilation) operator on atom $i$, and the summation runs over all adjacent atoms. The vector potential can be included in the Hamiltonian by the Peierls substitution of tight-binding version
\begin{eqnarray}\label{Piels}
\gamma \rightarrow \gamma e^{-i \boldsymbol{\mathcal{A}}(t)\cdot \boldsymbol{r}_\alpha} \approx
\gamma[1-i\, \boldsymbol{\mathcal{A}}(t)\cdot \boldsymbol{r}_\alpha],
\end{eqnarray}
where $\boldsymbol{r}_\alpha$ means the position vector form one atom to an adjacent atom. The Peierls substitution and the $A\cdot p$ approximation are equivalent for the bulk graphene (See the Appendix). The former is general and the latter is only applicable for the  translational invariant systems in which $\bm p$ is a good quantum number. 

We first calculate the dispersion energies and the corresponding quantum states of un-irradiated graphene ribbons by choosing a translational unit cell (see Fig. \ref{ribbon}), which includes $N$ atoms supposedly. Second, we calculate the transition matrix $g$ using the obtained eigen states. (Because $k_y$ is not a good quantum number, for one longitudinal wavevector $k_x$, there are $N$ eigen states. One half of them are conduction band states and the other half are valence band states. The transition element $g$, which is a pure number previously, now have to be treated as a matrix of dimension $N/2\times N/2$). Third, we construct the ribbon version of the Hamiltonian in Eq. (\ref{H}) for each $k_x$, which is a matrix of the dimension $N\times N$. Solving the eigen problem and according to Eq. (\ref{EF}), we have the $N$ Floquet energies (the numbers of $E^F_+$ and of $E^F_-$ for one value of $k_x$ are both $N/2$). Finally, we introduce the time-averaged state density,\cite{resonant_3} which can be calculated in the Floquet space. The calculation detail will be presented in the Appendix, and here we only list the result without proven as
\begin{eqnarray}\label{rho_reduced}
\begin{split}
\rho_0
&=\sum_{E^F_+} \left\{ |a_F|^2 \, \delta(E-E^F_{+}) + |b_F|^2 \,\delta[E-(E^F_+-\omega)] \right\} \\
&+\sum_{E^F_-} \left\{ |b_F|^2\, \delta(E-E^F_-) + |a_F|^2\,\delta[E-(E^F_-+\omega)] \right\}.
\end{split}
\end{eqnarray}
The state density can be viewed as weighted dispersion of the system under irradiation. In deed, Eq. (\ref{rho_reduced}) can be understood intuitively without complicated arithmetics. According to Eq. (\ref{psiF}), the Floquet state $\psi^F_+$ consists of two parts with different weights. Correspondingly, the quantum states of the system, $e^{-iE^F_+t}\psi^F_+$, is composed by $|c\rangle$ weighted by $|a_F|^2$ with the quasi energy $E^F_+$ and $|v\rangle$ weighted by $|b_F|^2$ with the quasi energy $E^F_+-\omega$, which results in the states density indicated in the first line of Eq. (\ref{rho_reduced}). Similar analysis of state $e^{-iE^F_-t}\psi^F_-$ leads to the state density in the second line.

Figure \ref{dispersion} shows the weighted dispersions of zigzag and armchair graphene ribbons under light irradiation with various polarization configurations. These dispersions are something like front-view of bulk dispersions showed in Fig. \ref{dos} a-c. It is interesting that one can observe Floquet edge bands within the resonant gaps for both zigzag and armchair ribbons under both linearly and circularly polarized illumination. For the linear polarization along $x$-direction, flat edge bands can be found on the zigzag ribbon but cannot on the armchair ribbon. While for the $y$-polarized irradiation, edge bands only arise on the armchair ribbon, and the edge bands are not flat but curved ones. For the case of circular polarization, edge bands can be found for both zigzag and armchair ribbons and two band-crossings take place in the resonant gaps at $E=\pm\omega/2$, reproducing the main results in Refs. [\onlinecite{resonant_4}]. If multi-photon processes are taken into account, as did in the reference, the edge bands crossings can be found indeed anti-crossings with the min-gap of $\sim|g|^3$, which is caused by the third-photon process and is extreme small to detect. The edge bands for circular polarization are helical (the quantum states of them go forward on one edge and comeback on the opposite edge), which are topologically different from edge bands for linear polarization. The topologic origination of the helical edge bands will be discussed later. All the edge bands, for both linear and circular polarizations, are nearly half-weighted. 

The dispersion properties are inherited from Eq. (\ref{H}).
Figure \ref{effective} shows the dispersion relations by solving the Hamiltonian in Eq. (\ref{H}) for zigzag and armchair graphene ribbons . The in-gap electronic structures are similar to those in Fig. \ref{dispersion}. Edge bands can be found for $x$-polarized illumination on zigzag ribbon, for $y$-polarized case on armchair ribbon, and for circular polarization on both types of ribbons. We investigate the wavefunctions of the edge states for the linear polarization cases, and verify that they are localized at ribbon edges and decay into the bulk, as Inset 1 and 2 demonstrate. The curved edge bands close to $E=0$ in Fig. \ref{effective} (d) are not very clear because of the interference of other states nearby, which can be improved by doing the calculation on wider ribbons, as Inset 3 shows. The wavefunctions of edge bands for the circular polarization case were extensively studied in Refs.[\onlinecite{resonant_3}] and [\onlinecite{resonant_4}], and so we skip over detailed discussion on them.

\section{ Chern number and winding number}

The edge bands for circular polarization case have their topologic origin, which can be digged out from Eq. (\ref{H}). The static Hamiltonian in the equation has the form of $\mathcal{H} =\boldsymbol{h}(\boldsymbol{k})\cdot\boldsymbol{\sigma}$.
When $\boldsymbol{k}$ changes through out the whole Brillouin zone, the endpoint of vector $\boldsymbol{h}$ maps out a closed surface, which  does or does not contain the origin. The Chern number of the upper band is defined by the number of times of the origin contained by the surface. If we cut the surface using the plane $h_z=0$, i.e., $\delta=0$, the intersection between the plane and the surface is a closed curve. The number of times of the origin contained by the surface is just that enclosed by the curve, In other words, the Chern number is reduced to the winding number at resonance. Setting $\delta=0$ and recalling $g=|g|e^{i\theta}$, the azimuthal angle of $\boldsymbol{h}=(h_x, h_y)$ is just the complex angle of $g$. So, the winding number reads
\begin{eqnarray}\label{winding}
\mathcal{C} = \frac1{2\pi}\oint_{\delta=0} d\theta.
\end{eqnarray}

\begin{figure}
\includegraphics[width=8cm]{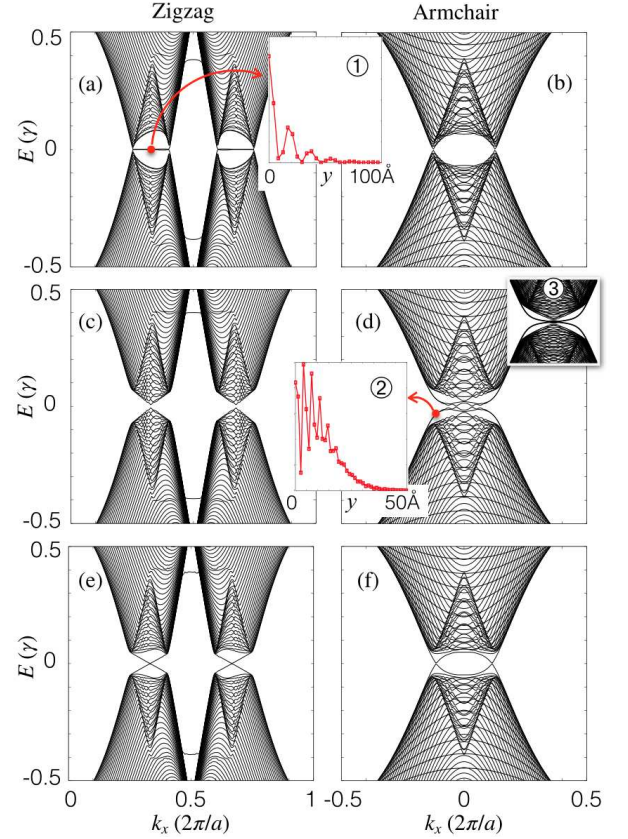}
\caption{(Color Online) \label{effective}
Dispersions obtained by solving the Hamiltonian in Eq. (\ref{H}) for graphene ribbons with 100 lateral atoms under (a-b) $x$-polarized irradiation with $A_x=0.1\gamma$, (c-d) $y$-polarized irradiation with $A_y=0.1\gamma$, and (e-f) circularly polarized irradiation with $\sqrt{|A_x|^2+|A_y|^2}=0.1\gamma$. Insets 1 and 2 show the wavefunction maglitude square on specified atom lines indicated by filled circles in Fig. \ref{ribbon}. Inset 3 is the zoom picture of (d) around the resonant gap of a wider ribbon with 300 lateral atoms. The frequency of the irradiation is $\omega=0.8\gamma$.}
\end{figure}

For linearly polarized light illuminating case, saying, $A_x=0$ or $A_y=0$ in Eq. (\ref{gK}), $g$ is a pure imaginary number. Under the constriction $\delta=0$, $\boldsymbol{k}$ evolutes as a circle with the radius $k=\omega/2$, Letting the azimuthal angle $\varphi$ change from 0 to $2\pi$, the endpoint of complex vector $g$ evolutes as a vertical line on the complex plane, as indicated in Figs. \ref{dos}. (g) and (h). Because the line lies across the origin, the upper part of the dispersion of the Hamiltonian in Eq. (\ref{H}) and the lower part have degenerate points and no real gap is formed (see Fig \ref{effective}). For the circular polarization, we have $g = Ae^{i\varphi}$. The endpoint of $g$ is a circle with the radius $A$ when $\varphi$ change from 0 to $2\pi$ and the winding number is 1. Figure \ref{dos} (i) shows the winding number picture for this situation.

For the other valley $K'$, the bulk Hamiltonian in Eq. (\ref{HK}) should be $H_{\boldsymbol{k}} = \boldsymbol{\sigma}^* \cdot \boldsymbol{k}$, and In Eqs. (\ref{eigenK}) and (\ref{gK}) $\varphi$ is replaced with $-\varphi$. For the circularly polarized irradiation case, we have the transition element $g = Ae^{-i\varphi}$. Because of the time-reversal symmetry between the two valleys, $\boldsymbol{k}$ evolutes reversely with respect to that in valley $K$ (from $2\pi$ to 0), $g$ as a complex vector varies in the same way as before, and the winding number for valley $K'$ is 1 too. In total, the Chern number of the system described by Eq. (\ref{H}) is 2, which implies there are two crossings of edge bands (four edge bands) regardless of the edge type of graphene ribbon and the edge bands are helical ones: the forward moving states of the edge bands are localized on one edge and backward on the other edge. The helicity of the edge bands are controlled by the rotation of the circular polarization. If reversing the polarization, saying, $A_x=A$ and $A_y=-iA$, the evolution of the endpoint of $g$ is reversed, the Chern number turns to be $-2$, and the states of the edge bands flow reversely.

For the circular polarization case, the two crossings and four edge bands in the gap at $E=0$ in Fig. \ref{effective} are shared by two resonant gaps in the Floquet spectrum at $E=\pm \omega/2$ in Fig. \ref{dispersion}, i.e., two edge bands and one crossing in one gap. For the zigzag ribbon, the two valleys are resolved and the two edge bands in the gap $E=\omega/2$ (or $-\omega/2$) are equally hosted by the two valleys, saying, one edge band in one valley. The edge band is composed of a half-weighted forward-flow band and a half-weighted backward-flow band. 

\section{Summary}
A theoretical model was proposed to describe the Floquet band structure of graphene under irradiation by taking into resonant inter-band transition. Linearly polarized light induces pseudo gaps, and circularly polarized light causes real gaps on the quasi-energy spectrum. If the polarization of light is linear and along the longitudinal direction of zigzag ribbons, flat edge bands appear in the pseudo gaps, and if is in the lateral direction of armchair ribbons, curved edge bands can be found. For the circularly polarized cases, edge bands arise and intersect in the gaps of both types of ribbons. The edge bands induced by the circularly polarized light are helical and those by linearly polarized light are topologically trivial ones. We reduced the Chern number of the Floquet band into the winding number at resonance.

\acknowledgements

This work was supported by NSF of China Grant No. 11274124, 
No. 11474106, No. 11174088, and No. 11575051.

\appendix

\section{Connection between $A \cdot p$ approximation and Peierls substitution}
We have used the $A \cdot p$ approximation in analytical analysis and the Peierls substitution of tight binding version for numerical calculation on graphene ribbons. For the bulk graphene, the Peierls substitution will result in the $A \cdot p$ approximation.

In basis of $A$-$B$ sublattices, the $k$-space Hamiltonian of bulk graphene reads 
\begin{eqnarray} \label{H_App}
H_{\bm k} = 
\left[\begin{array}{*{20}c}
0 & h_{\bm k} \\
h_{\bm k}^* & 0
\end{array}\right] \quad {\rm with}\quad  h_{\bm k}=
\sum_\alpha \gamma e^{i{\bm k}\cdot {\bm r}_\alpha}, 
\end{eqnarray}
where $\boldsymbol{r}_\alpha$ are the position vectors from an $A$-atom pointing to adjacent $B$-atoms. When the irradiation is present, the Peierls substitution in Eq. (\ref{Piels}) results in 
\begin{eqnarray}
\begin{split}
h_{\bm k} &\rightarrow
\sum_\alpha \gamma[1-i\, \boldsymbol{\mathcal{A}}\cdot \boldsymbol{r}_\alpha] e^{i\boldsymbol{k}\cdot \boldsymbol{r}_\alpha} \\
&=\; h_{\boldsymbol{k}} - i \boldsymbol{\mathcal{A}}\cdot  \sum_\alpha  \gamma\boldsymbol{r}_\alpha  e^{i\boldsymbol{k}\cdot \boldsymbol{r}_\alpha}\\
&=\; h_{\boldsymbol{k}} - \boldsymbol{\mathcal{A}}\cdot \nabla_{\boldsymbol{k}}h_{\boldsymbol{k}}.
\end{split}
\end{eqnarray}
This leads to the change of the Hamiltonian,
\begin{eqnarray}
\begin{split}
H_{\boldsymbol{k}} &\rightarrow H_{\boldsymbol{k}} - \boldsymbol{\mathcal{A}}\cdot \nabla_{\boldsymbol{k}}H_{\boldsymbol{k}} \\
&= \; H_{\boldsymbol{k}} - \boldsymbol{\mathcal{A}}\cdot \boldsymbol{p},
\end{split}
\end{eqnarray}
in which the identity ${\bm p}= \nabla_{\boldsymbol{k}}H_{\boldsymbol{k}} $ is used. On can see the additional term caused by the irradiation is just the perturbation of the $A\cdot p$ approximation.


\section{State density}
The Floquet Hamiltonian can be expressed as a static matrix in the Floquet space, the basis of which are chosen as $\left\{(|v\rangle, |c\rangle )e^{in\omega t}\right\}$. Because only one-photon processes are considered, the basis are truncated to be 
\begin{eqnarray}
(|v\rangle, |c\rangle )e^{i\omega t}, \;\;\; (|v\rangle, |c\rangle ), \;\;\; (|v\rangle, |c\rangle )e^{-i\omega t}. \nonumber
\end{eqnarray}
The matrix element of an arbitrary operator $\hat{O}$ between the states $|\mu\rangle e^{-im\omega t}$ and $|\nu\rangle e^{-in\omega t}$ is defined by
\begin{eqnarray}
\frac{1}{T} \int_0^T  \langle \mu |\hat{O}| \nu \rangle e^{i(m-n) \omega t} dt, \nonumber
\end{eqnarray}
where $T=2\pi/\omega$ is the period of the time-dependent parameter. The Floquet Hamiltonian in the Floquet space reads
\begin{eqnarray}
H^F &= \left(\begin{array}{*{20}c}
\epsilon_v-\omega & 0 &   &   &   &  \\
0 & \epsilon_c-\omega & g^*/2 &   &   &  \\
  & g/2 & \epsilon_v & 0 &   &  \\
  &   & 0 & \epsilon_c & g^*/2 &  \\
  &   &   & g/2 & \epsilon_v+\omega & 0  \\
  &   &   &   & 0 & \epsilon_c+\omega
\end{array}\right). \nonumber
\end{eqnarray}
The matrix is is block diagonal and we rewrite the it block-by-block as
\begin{eqnarray}
H^F &= 
\left(\begin{array}{*{20}c} 
\epsilon_v-\omega &   &  &  \\
  & \mathcal{H}+(\epsilon_A-\omega)/2 &   &   \\
  &   & \mathcal{H}+(\epsilon_A+\omega)/2 &    \\
  &   &   & \epsilon_c+\omega
\end{array}\right), \nonumber
\end{eqnarray}
where $\epsilon_A=(\epsilon_c+\epsilon_v)/2$ is the average band energy  and $\mathcal{H}$ is the Hamiltonian matrix defined in Eq. (\ref{H}). The Floquet Green's function $G^F(E)=(E-H^F)^{-1}$ can be easily calculated by matrix inversion block-by-block. The spectrum operator in the Floquet space is defined as 
\begin{eqnarray}
\rho = \frac1 {2\pi} [G^F(E+i0^+)-G^F(E-i0^+)].\nonumber
\end{eqnarray}
The time-averaged state density can be obtained by tracing the spectrum matrix in the zero-photon subspace,\cite{resonant_3} saying.
\begin{eqnarray}
\rho_0 = \sum_{n=3,4} \rho_{n,n}.\nonumber
\end{eqnarray}
The calculation is straight and the result reads
\begin{eqnarray}
\begin{split}
\rho_0
&= |a_F|^2 \, \delta(E-E^F_{+}) + |b_F|^2 \,\delta[E-(E^F_+-\omega)]   \\
&+ |b_F|^2\, \delta(E-E^F_-) + |a_F|^2\,\delta[E-(E^F_-+\omega)].
\end{split} \nonumber
\end{eqnarray}
The equation is derived for bulk graphene, and the ribbon version of it is just Eq. (\ref{rho_reduced}).


\begin{references}
\bibitem{explain} In Eq.(\ref{Ht_original}), the $A\cdot p$ term is $\boldsymbol{\mathcal{A}} \cdot \boldsymbol{p}/m$ ($m$ is set to be 1 so it is invisible), The corresponding term for graphene is not that by setting $m\rightarrow 0$, but just simply get rid of $m$.
\end{references}
\end{document}